\documentclass[prl, twocolumn, superscriptaddress]{revtex4}
\usepackage{graphicx,color}
\usepackage{amssymb}
\usepackage{amsmath}
\usepackage{bm}
\usepackage{hyperref}
\usepackage{tabularx}
\usepackage{multirow}
\usepackage{braket}
\usepackage{xcolor}
\usepackage{gensymb}

\begin{document}

\title{Electronic Density of States of a $U\left(1\right)$ Quantum Spin Liquid with Spinon Fermi Surface. II. Zeeman Magnetic Field Effects}
\author{Wen-Yu He}\thanks{hewy@shanghaitech.edu.cn}
\affiliation{School of Physical Science and Technology, ShanghaiTech University, Shanghai 201210, China}
\author{Patrick A. Lee}\thanks{palee@mit.edu}
\affiliation{Department of Physics, Massachusetts Institute of Technology, Cambridge, Massachusetts 02139, USA}

\date{\today}

\begin{abstract}
The Zeeman effect lowers the energy of electrons with spin states which are  anti-parallel to the applied magnetic field but lifts that of spin parallel states. In quantum spin liquids where the spin and charge degrees of freedom are fractionalized,  anomalous Zeeman response may be expected. In the case of spin liquids with spinon Fermi surface, the threshold energy to excite an electronic state is found to exhibit no Zeeman shift. This is specific to  the spinon Fermi surface case. In contrast, other gapped spin liquids are expected to exhibit the standard Zeeman shift at the band edge even though they also exhibit spin-charge fractionalization. When gauge field fluctuations are included, we find that the Zeeman shift of the electronic states gets affected by the gauge field induced binding. In the electronic density of states spectra, weak gauge binding induces band edge resonance peaks which exhibit the Zeeman shift in the same direction as that in the standard Zeeman effect, but the shift is reduced as the  binding potential increases. With further increase in the binding potential the resonance becomes true in-gap bound states and eventually the Zeeman shift direction reverses so it is  opposite to  the standard Zeeman effect. We propose that one can perform  spin polarized scanning tunneling microscope measurements as a test of the spinon Fermi sea ground state in  quantum spin liquid candidate materials.
\end{abstract}

\maketitle

\section{Introduction}
The Zeeman response of an electronic state to an external magnetic field is well understood since the establishment of quantum mechanics~\cite{Zeeman1, Zeeman2, Griffths1}. In the standard Zeeman effect, the Zeeman coupling between the electron's spin and an external magnetic field shifts the electronic energy~\cite{Griffths1, Griffths2} in a direction which depends on the spin orientation: electrons with spins parallel to the magnetic field have the energy shifting up while those with spins anti-parallel have the energy shifting down. This standard picture assumes that the electron carries both the charge and spin.

In modern condensed matter physics, it is known that strong electronic correlations can give rise to exotic states of matter such as quantum spin liquid (QSL)~\cite{Lee1,Savary,YiZhou}, where the spin degree gets deconfined~\cite{XiaoGang, SungSicLee}. In a QSL, a physical electron undergoes spin-charge separation and is fractionalized into a chargon and a spinon~\cite{Florens, Lee2}. A chargon is a charged boson that carries the electric charge, while a spinon is a charge neutral fermion that carries spin-1/2. One particularly important QSL is the QSL with spinon Fermi surface (SFS)~\cite{Lee1, Lee2, Senthil}. The QSL with SFS is a charge insulator but has gapless itinerant spinon excitations from the SFS. In the QSL with SFS, since an electron is a composite particle with the constituent spinon and chargon spatially separated, it becomes intriguing to see how the electronic Zeeman response in the QSL with SFS can differ from the response in the standard Zeeman effect. 

In this work, we show that the electrons in the QSL with SFS exhibit anomalous Zeeman response in three aspects: 1) the Zeeman field has no effect on the threshold energy to excite an electron or a hole in the QSL; 2) the gauge field fluctuations reduce the Zeeman shift of the quasi-bound spinon chargon pairs; 3) strong gauge field fluctuations can induce in-gap bound state~\cite{Lee3, Lee4} in the QSL and the Zeeman field shifts the bound state energy in the direction opposite to that in the standard Zeeman effect. In the QSL, the Zeeman field only couples to the spinons, but the electronic Zeeman response requires the consideration of the {constituent} spinon and chargon as a whole. We will show that such internal composite structure of electrons along with the spinon Fermi sea ground state in the QSL gives rise to the anomalous electronic Zeeman response.

The QSL phase is one of the most fascinating states in the condensed matter due to its potential to host various exotic quantum phemonenon~\cite{Lee1,Savary,YiZhou, Anderson1, Kitaev, Nayak}. Recently, a series of Mott insulators~\cite{Saito, Khaliullin, McQueen, Plumb, HYKee, Qingming, Law, Law2} have been proposed as the candidates to host the QSL phase, but conclusive evidence has yet been arrived in spite of the intensive experimental investigations~\cite{YSLee, Mori, JunZhao, Nagler, SYLi, Matsuda, Ong}. Here, based on the newly discovered anomalous electronic Zeeman response in the QSL with SFS, we propose to use a spin polarized scanning tunneling microscope (STM) to detect the spinon Fermi sea ground state in the QSL candidate materials. In the electronic DOS spectra, we emphasize that the absence of Zeeman shift of the threshold energy provides strong evidence for the existence of the SFS.

\begin{figure}
\centering
\includegraphics[width=3.5in]{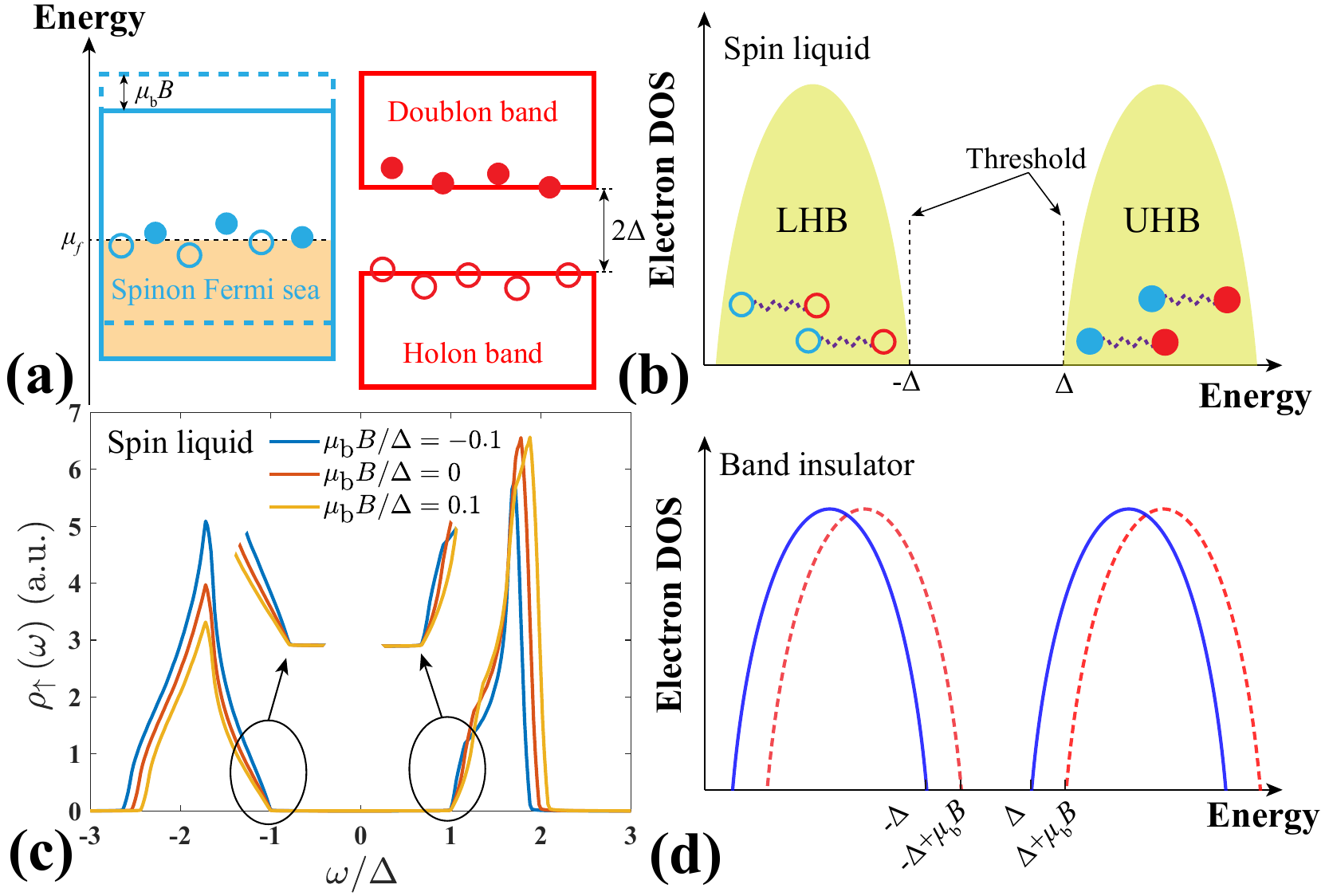}
\caption{(a) The schematic showing of the spinon, holon and doublon bands in the QSL. The shaded region in the spinon band represents the Fermi sea filled by the spinons. The filled blue and red circles denote the spinons and doublons respectively, while the empty blue and red circles are the 
{antispinons} and holons respectively. The dashed blue rectangle demonstrates the Zeeman field induced energy shift of the spinon band. (b) The schematic plot of the QSL electronic DOS spectra. The antispinon holon pairs and the spinon doublon pairs are the states in the LHB and UHB respectively. The spinons and chargons are coupled through the emergent $U\left(1\right)$ gauge field that is represented by the purple wavy lines. (c) The spin up electronic DOS calculated in a triangular lattice given different Zeeman fields. The Zeeman field has no effect on the threshold but changes the Hubbard band widths. (d) The electronic DOS of a band insulator. The blue lines denote the case of $B=0$T while the red dashed lines show the standard energy shift of $\mu_{\textrm{b}}B$ induced by a finite Zeeman field {applied on the spin up electrons.} {Here $\omega$ denotes the energy and $\Delta$ is the charge gap.}}\label{figure1}
\end{figure}

\section{Absence of Zeeman shift in the threshold energy}
Our analysis of the electronic Zeeman effect in the QSL with SFS starts from the slave rotor mean field Hamiltonian of the QSL~\cite{Supp}:
\begin{align}
H_0=\sum_{\bm{k}}\epsilon_{\bm{k}}\left(a_{-\bm{k}}a^\dagger_{-\bm{k}}+b^\dagger_{\bm{k}}b_{\bm{k}}\right)+\sum_{\bm{k},\sigma}E_{\sigma, \bm{k}}f^\dagger_{\sigma,\bm{k}}f_{\sigma,\bm{k}},
\end{align}
where $a^{\left(\dagger\right)}_{-\bm{k}}$, $b^{\left(\dagger\right)}_{\bm{k}}$ and $f^{\left(\dagger\right)}_{\sigma,\bm{k}}$ are the annihilation (creation) operators for a holon, doublon and spinon respectively. Here the holons and doublons carry the charge $\pm e$ respectively and they are the nonrelativistic approximation to the chargons~\cite{Lee2}. The chargon band is $\epsilon_{\bm{k}}$. The index $\sigma=\uparrow/\downarrow$ denotes the spin. The spinon bands are $E_{\sigma,\bm{k}}$. Taking into account the spinon chemical potential $\mu_f$, we define $\xi_{\sigma,\bm{k}}=E_{\sigma,\bm{k}}-\mu_f$. Since $\mu_f$ does not change with magnetic field to linear order, a finite magnetic field $\bm{B}=\left(0,0,B\right)$ changes the spinon band dispersions to $\xi_{\uparrow/\downarrow,\bm{k}}=\xi_{\bm{k}}\pm\frac{1}{2}g\mu_{\textrm{b}}B$, where $g=2$ is the Land\'e $g$ factor and $\mu_{\textrm{b}}$ is the Bohr magneton. In the QSL with SFS, the spinon chemical potential $\mu_f$ lies inside the spinon bands, so the spinon states below $\mu_f$ all get filled as indicated in Fig. \ref{figure1} (a). The resulting ground state is the Fermi sea of spinons: $\ket{\textrm{G}}=\prod_{\bm{k},\xi_{\sigma,\bm{k}}\leqslant0}f^\dagger_{\uparrow,\bm{k}}f^\dagger_{\downarrow,\bm{k}}\ket{0}$. The chargon band has a gap $\Delta=\textrm{min}\left[\epsilon_{\bm{k}}\right]$, so the holon branch and the doublon branch are separated in energy as can be seen in Fig. \ref{figure1} (a).

In the QSL, annihilating a spinon and creating a holon results in the excitation of a hole state, while creating a spinon along with a doublon leads to the creation of an electron. At zero temperature, the energy cost to excite a hole is $\bra{\textrm{G}}a_{-\bm{k}'}f^\dagger_{\sigma,\bm{k}}\left(H_0-\mu_f\hat{N}_f\right)f_{\sigma,\bm{k}}a^\dagger_{-\bm{k}'}\ket{\textrm{G}}-\bra{\textrm{G}}H_0-\mu_f\hat{N}_f\ket{\textrm{G}}=-\xi_{\sigma,\bm{k}}+\epsilon_{\bm{k}'}$ with $\xi_{\sigma,\bm{k}}\leqslant0$. Similarly, the energy cost to excite an electron is $\bra{\textrm{G}}b_{\bm{k}'}f_{\sigma,\bm{k}}\left(H_0-\mu_f\hat{N}_f\right)f_{\sigma,\bm{k}}^\dagger b^\dagger_{\bm{k}'}\ket{\textrm{G}}-\bra{\textrm{G}}H_0-\mu_f\hat{N}_f\ket{\textrm{G}}=\xi_{\sigma,\bm{k}}+\epsilon_{\bm{k}'}$ with $\xi_{\sigma,\bm{k}}\geqslant0$. Here $\hat{N}_f=\sum_{\sigma,\bm{k}}f^\dagger_{\sigma,\bm{k}}f_{\sigma,\bm{k}}$ is the spinon number operator. It is clear that the minimum energy cost to excite an electron or a hole in the QSL equals to the charge gap $\Delta$ because the minimum spinon excitation is at the Fermi momentum and cost zero energy. The minimum energy cost to excite an electron or a hole in the QSL is defined as the threshold energy. 

In a finite magnetic field, the Zeeman coupling lifts the spinon band of parallel spin as schematically shown in Fig. \ref{figure1} (a), while the spinon band of anti-parallel spin gets lowered. In the region of $\mu_{\textrm{b}}B<\mu_f$, the Zeeman field imbalances the spinon Fermi sea occupation of the two spin species. The Fermi surface areas of the spin parallel and anti-parallel electrons become unequal but the chemical potential is unchanged. As long as the Fermi surfaces of the two spin species exist inside the charge gap, the threshold energy always equals to $\Delta$ regardless of the spin orientations.

In the energy range beyond the threshold energy, excitations of antispinon holon pairs and spinon doublon pairs generate the continuum spectrum of the lower Hubbard band (LHB) and the upper Hubbard band (UHB) respectively in the electronic DOS spectra, which is schematically shown in Fig. \ref{figure1} (b). The LHB top and the UHB bottom denote the threshold to create a physical electronic excitation. Since the Zeeman field does not change the threshold energy, no shift is expected for the threshold in the QSL electronic DOS spectra.

To verify the absence of Zeeman shift in the threshold energy analyzed above, we perform a calculation on the QSL electronic DOS in a triangular lattice. The band dispersions of the spinon band and the chargon band are taken to be $\xi_{\bm{k}}=-2t_f\left(2\cos\frac{1}{2}k_xa\cos\frac{\sqrt{3}}{2}k_ya+\cos k_xa\right)-\mu_f$ and $\epsilon_{\bm{k}}=-2t_X\left(2\cos\frac{1}{2}k_xa\cos\frac{\sqrt{3}}{2}k_ya+\cos k_xa-3\right)+\Delta$ respectively. Here $a$ is the lattice constant and the band parameters are set to be $t_f=0.03$ eV, $t_X=0.02$ eV, $\Delta=0.25$ eV and $\mu_f=0.025$ eV. The QSL electronic DOS is given by $\rho_\sigma\left(\omega\right)=-\frac{1}{\Omega N\pi}\sum_{\bm{k},\bm{k}'}\textrm{Im}G_\sigma^{\textrm{R}}\left(\omega,\bm{k},\bm{k}'\right)$, where $\Omega$ denotes the sample area and $N$ is the number of lattice sites. Here the electronic retarded Green's function $G_\sigma^{\textrm{R}}\left(\omega,\bm{k},\bm{k}'\right)$ takes the form~\cite{Supp}
\begin{align}\label{GR01}
G_{\sigma}^{\textrm{R}}\left(\omega,\bm{k},\bm{k}'\right)=&\frac{n_{\textrm{F}}\left(\xi_{\sigma,\bm{k}}\right)+n_{\textrm{B}}\left(\epsilon_{\bm{k}'}\right)}{\omega+i0^+-\xi_{\sigma,\bm{k}}+\epsilon_{\bm{k}'}}+\frac{n_{\textrm{F}}\left(-\xi_{\sigma,\bm{k}}\right)+n_{\textrm{B}}\left(\epsilon_{\bm{k}'}\right)}{\omega+i0^+-\xi_{\sigma,\bm{k}}-\epsilon_{\bm{k}'}}
\end{align}
with $n_{\textrm{F}}\left(\xi\right)$ and $n_{\textrm{B}}\left(\epsilon\right)$ being the Fermi and Bose distribution functions respectively. The resulting spin up electronic DOS in different Zeeman fields is plotted in Fig. \ref{figure1} (c) and the case of spin down can be deduced by reversing the magnetic field direction. Consistent with our analysis, the threshold exhibits no shift in the Zeeman field. Interestingly, the LHB bottom and the UHB top are observed to shift by $\mu_{\textrm{b}}B$ in the same direction as that in the standard Zeeman effect. Since the states at the LHB bottom and the UHB top involve the excitations of an antispinon at the spinon band bottom and a spinon at the spinon band top respectively, the observed shift of the LHB bottom and the UHB top originates from the Zeeman shift of the spinon band. As a result, in the QSL with SFS, the Zeeman field has no effect on the threshold but changes the Hubbard band width. In sharp contrast, a band insulator has its threshold shifted with the Zeeman field while the band width does not change, as is illustrated in Fig. \ref{figure1} (d).

We note that in a gapped QSL, the spinon spectrum acquires a gap $\Delta_{\textrm{s}}$, so the threshold energy will be at $\Delta_{\textrm{s}}+\Delta$. In a Zeeman field, the energy gap in the spinon spectrum shifts in the standard way, so the threshold energy to excite an electron or a hole also exhibits the standard Zeeman shift even when there is spin charge separation and deconfinement in the gapped QSL. Thus the absence of the Zeeman shift in the threshold is a unique signature of the QSL with SFS.

\begin{figure}
\centering
\includegraphics[width=3.5in]{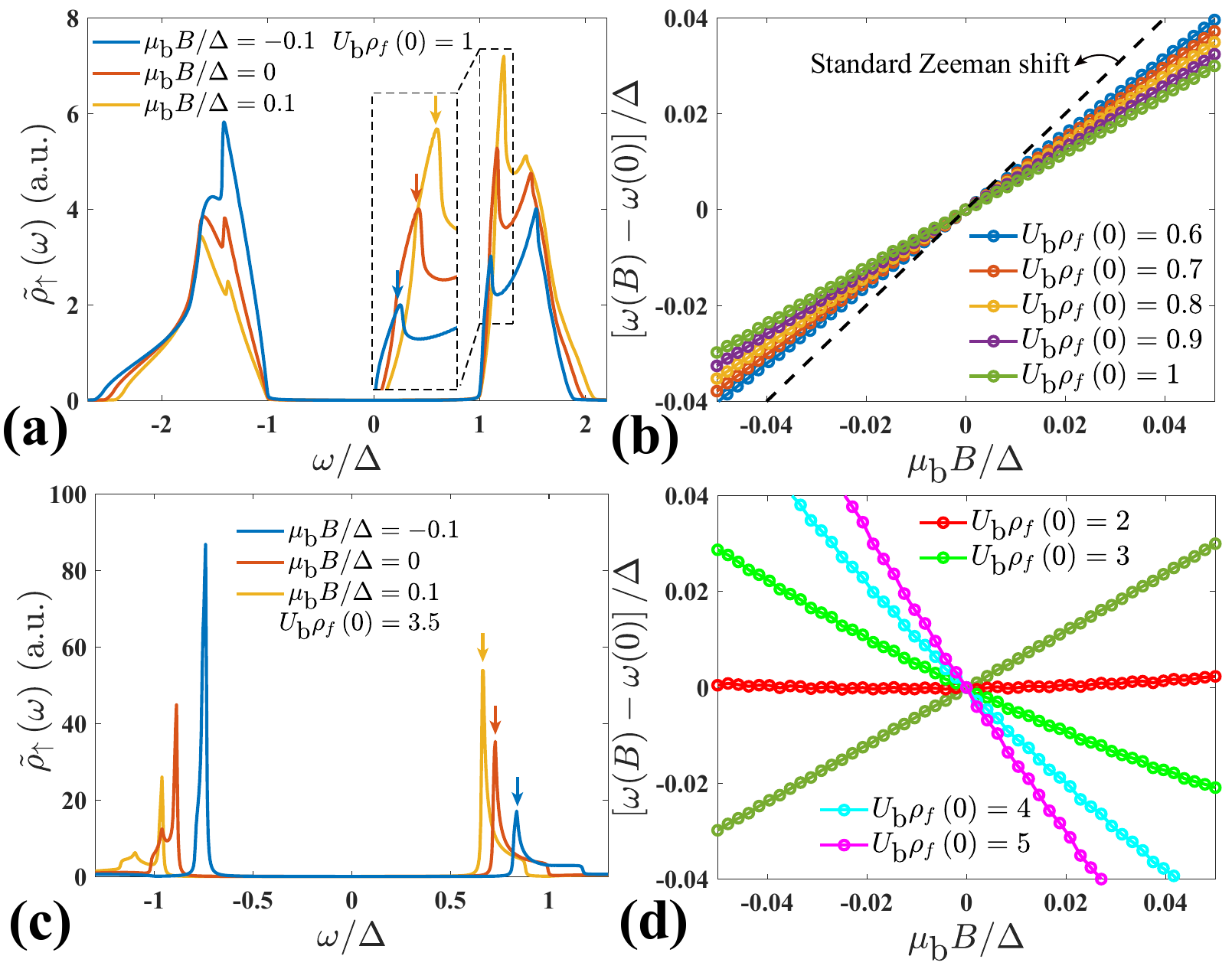}
\caption{(a) The spin up electronic DOS in the QSL {with} the gauge binding $U_{\textrm{b}}\rho_f\left(0\right)=1$. The gauge binding induces band edge resonance peaks and the resonance peaks exhibit the Zeeman shift in the same direction as that in the standard Zeeman effect. The inset is the zoom of the UHB edge resonance peaks. The threshold remains unchanged in the Zeeman field. (b) The linear dependence of the UHB edge resonance peak energy on the Zeeman field. The slope is found to decrease with the increase of the gauge binding interaction. The black dashed line denotes the energy shift in the standard Zeeman effect. (c) The in-gap bound state peaks in the QSL electronic DOS spectra. The gauge binding is $U_{\textrm{b}}\rho_f\left(0\right)=3.5$. In the Zeeman field, the in-gap bound state peaks move in the direction opposite to that in (a). (d) The reversal of the slope in (b) as the gauge binding interaction continues increasing. {Here $\omega$ denotes the energy, $\Delta$ is the charge gap, and $B$ denotes the magnetic field along the $z$ direction.}}\label{figure2}
\end{figure}

\section{Naive Zeeman shift of the quasi-bound spinon chargon pairs}
The zero Zeeman shift of the threshold in the electronic DOS spectra shown in Fig. \ref{figure1} (c) stems from the electron fractionalization and the spinon Fermi sea ground state in the QSL. In this case the fractionalized quasi-particles, the spinons and the chargons, both couple to an emergent $U\left(1\right)$ gauge field~\cite{Lee2, Senthil} and the gauge field fluctuations in turn affect the electronic DOS. It raises the question about how the electronic Zeeman response gets affected by the gauge field fluctuations in the QSL.

At the energy near the Hubbard band edges, the longitudinal component of the gauge field fluctuations plays the dominant role~\cite{Lee4} because the small {current-current} correlations there make the transverse components negligible. The longitudinal gauge field fluctuations bring about the gauge binding interaction~\cite{Lee4, Wenyu2}:
\begin{align}
H_{\textrm{int}}=&\frac{U_{\textrm{b}}}{N}\sum_{\sigma,\bm{k},\bm{q},\bm{q}'}f^\dagger_{\sigma,\bm{k}-\bm{q}}f_{\sigma,\bm{k}-\bm{q}'}\left(a_{-\bm{q}}a^\dagger_{-\bm{q}'}-b^\dagger_{\bm{q}}b_{\bm{q}'}\right).
\end{align}
Here $U_{\textrm{b}}$ denotes the gauge binding interaction strength which we assume to be short range and on-site due to the screening by spinons. An antispinon tends to bind a holon to form a hole bound state above the LHB, and similarly a spinon tends to bind a doublon to form an electronic bound state below the UHB. The spinon spinon repulsion from the spin fluctuations~\cite{Starykh} is not included as we focus on the QSL electronic state where the combination of a spinon and a chargon matters. Including the gauge binding, the QSL Hamiltonian becomes $H=H_0+H_{\textrm{int}}$ and the electronic DOS reads~\cite{Supp}
\begin{align}\nonumber
\tilde{\rho}_\sigma\left(\omega\right)=&-\frac{1}{\Omega\pi}\sum_{\bm{k}}\textrm{Im}\frac{\frac{1}{N}\sum_{\bm{q}}G^{\textrm{R}}_{h,\sigma}\left(\omega,\bm{k}-\bm{q},\bm{q}\right)}{1-\frac{U_{\textrm{b}}}{N}\sum_{\bm{q}}G^{\textrm{R}}_{h,\sigma}\left(\omega,\bm{k}-\bm{q},\bm{q}\right)}\\
&-\frac{1}{\Omega\pi}\sum_{\bm{k}}\textrm{Im}\frac{\frac{1}{N}\sum_{\bm{q}}G^{\textrm{R}}_{d,\sigma}\left(\omega,\bm{k}-\bm{q},\bm{q}\right)}{1+\frac{U_{\textrm{b}}}{N}\sum_{\bm{q}}G^{\textrm{R}}_{d,\sigma}\left(\omega,\bm{k}-\bm{q},\bm{q}\right)},
\end{align}
with $G^{\textrm{R}}_{h,\sigma}\left(\omega,\bm{k},\bm{k}'\right)$ and $G^{\textrm{R}}_{d,\sigma}\left(\omega,\bm{k},\bm{k}'\right)$ being the first and second terms in Eq. \ref{GR01} respectively.

In a moderate gauge binding of $U_{\textrm{b}}\rho_f\left(0\right)=1$, the QSL electronic DOS is found to have a pair of resonance peaks emerging at the Hubbard band edges as shown in Fig. \ref{figure2} (a). Here the $\rho_f\left(0\right)$ is the spinon DOS at the Fermi energy multiplied by $\Omega/N$. The rise of the band edge resonance peaks is interpreted as a pile-up of spectral weight to the Hubbard band edges induced by the quasi-bound spinon chargon pairs in relatively weak gauge binding~\cite{Wenyu2, Yichen}. Importantly, the threshold in the electronic DOS spectra in Fig. \ref{figure2} (a) also exhibits no Zeeman shift as that in Fig. \ref{figure1} (c), but the Zeeman field shifts the band edge resonance peaks in Fig. \ref{figure2} (a) in the same direction as that in the standard Zeeman effect. The shift inherits from the Zeeman field induced energy shift of the spinon in the quasi-bound spinon chargon pairs and resembles the standard Zeeman shift, so it is referred to the naive Zeeman shift. Such naive Zeeman shift applies generally to the resonance peaks formed from the spinon chargon pairs inside the Hubbard bands. In the weak gauge binding regime, the spinon chargon pairs become quasi-bound, leading to the band edge resonance peaks in the electronic DOS spectra. In this case the naive Zeeman shift becomes recognizable. Importantly, the naive Zeeman shift is found to reduce in magnitude from the standard Zeeman shift, and the reduction gets stronger as the gauge binding interaction increases. In Fig. \ref{figure2} (b), the energy of the UHB edge resonance peak as a function of the Zeeman field is plotted given different gauge binding interactions. The energy shift displays the expected linear dependence on $\mu_{\textrm{b}}B$, but the slope decreases from 1 as the gauge binding interaction increases. The deviation indicates that even in the naive Zeeman shift, the gauge binding gradually weakens the role of spinon Zeeman shift in determining the total energy of a spinon chargon pair.

\begin{figure}
\centering
\includegraphics[width=3.5in]{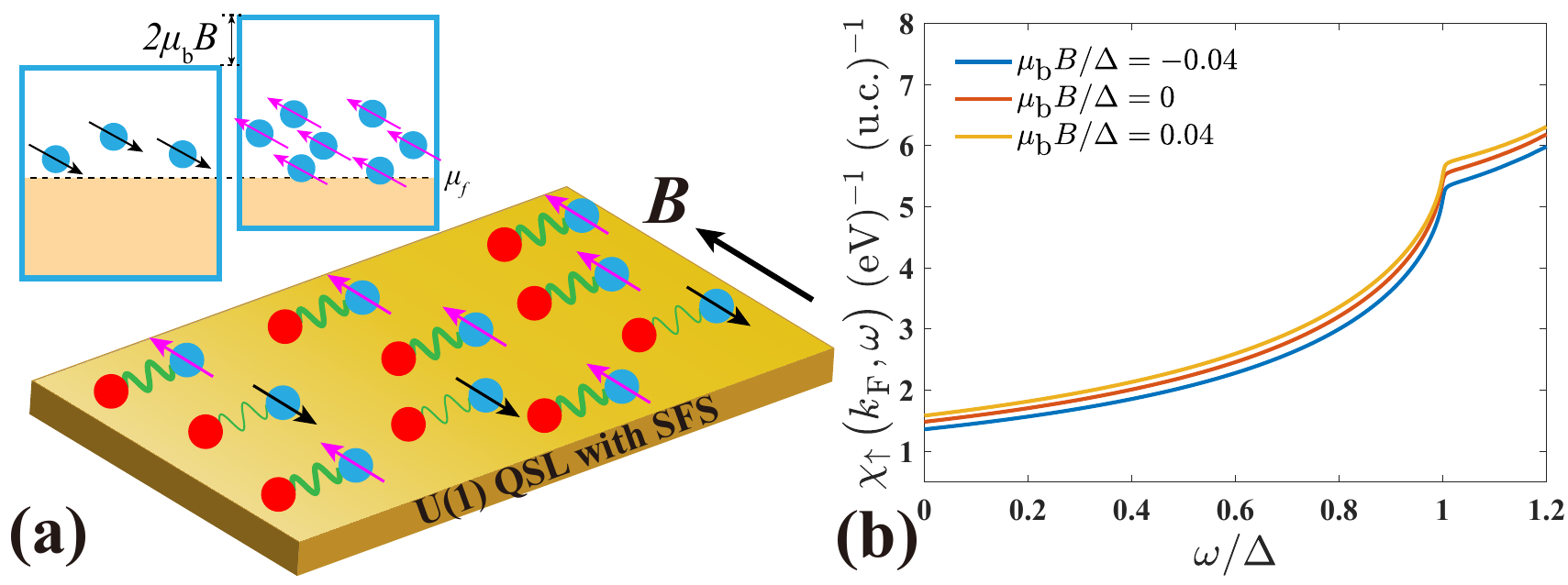}
\caption{(a) A schematic illustration of the Zeeman field effect on the binding. {The upper left panel shows that in the Zeeman field, more spinon states with parallel spin are available for injection than those of anti-parallel spin (The black and the purple arrows represent the anti-parallel and parallel spins respectively).} Since the Zeeman field increases the number of spinon empty states with parallel spin, the binding energy in the parallel spin species increases. Given a sufficiently large $U_{\textrm{b}}$, the energy of the bound state with parallel spin gets lowered by the Zeeman field. The green wavy line denotes the binding between a spinon and a chargon. The arrows in the spinons denote the spin orientations in the magnetic field. Here the spin quantization axis is set to be along the in-plane direction. (b) The susceptibility to bind a spin up spinon at the Fermi surface with a doublon in different Zeeman fields. The notation u.c. is short for unit cell. {Here $\omega$, $\Delta$ and $B$ denote the energy, charge gap and the magnetic field along the $z$ direction respectively.}}\label{figure3}
\end{figure}

\section{Anomalous Zeeman shift of the bound states}
Since the Zeeman shift of the band edge resonance peaks is observed to get suppressed by the increasing gauge binding in the range of $0.6\leqslant U_{\textrm{b}}\rho_f\left(0\right)\leqslant 1$, it becomes interesting to check how is the Zeeman shift evolving given a sufficiently large gauge binding. Surprisingly, in a strong gauge binding of $U_{\textrm{b}}\rho_f\left(0\right)=3.5$, the gauge binding induced peaks in Fig. \ref{figure2} (c) are observed to exhibit the Zeeman shift in the opposite direction compared to that in the standard Zeeman effect. In Fig. \ref{figure2} (d), the evolution of the UHB edge peak with the Zeeman field has its slope continuously decreases to 0, reverses sign and continue to increase in magnitude as the gauge binding increases in the range of $1\leqslant U_{\textrm{b}}\rho_f\left(0\right) \leqslant 5$. In the strong gauge binding regime, such Zeeman shift opposite to that in the standard Zeeman effect is referred to the anomalous Zeeman shift.

The anomalous Zeeman shift in the strong gauge binding regime is due to the interplay between the Zeeman field and the gauge binding induced energy saving, which is schematically shown in Fig. \ref{figure3} (a). Given a sufficiently large gauge binding, spinons and chargons are bound to form in-gap bound states~\cite{Wenyu2, Yichen, Wenyu3}. The in-gap bound states are manifested as the in-gap peaks in the electronic DOS spectra shown in Fig. \ref{figure2} (c). In the Zeeman field, due to the Pauli paramagnetism, the spinon states with spins parallel to the magnetic field are shifted up in energy, and there are more empty states available to inject a spinon. This increases the binding energy, resulting in a downward shift of the bound state energy. Note that this works only if a large fraction of the unoccupied states up to the upper band edge participate in binding. It requires a strong binding interaction comparable to the bandwidth, which is indeed the case when the bound state is split off below threshold.

%resless occupied than those with parallel spins, indicating that there are more available spinon excitations with anti-parallel spins. Intuitively, as the number of spinon excitations with anti-parallel spins increases, it becomes easier to squeeze one spinon excitation out of the crowd to get bound with a doublon. As a result, the binding of a doublon and a spinon of anti-parallel spin in the Zeeman field saves more energy than that in the case of $B=0$T, and it corresponds to the Zeeman field induced shift of the DOS peak in the direction of decreasing energy. Accordingly, the decrease in the number of spinon excitations with parallel-spins reduces the energy saving in the binding, so the Zeeman field shifts the corresponding DOS peak in the direction of increasing energy. The above analysis of the spinon doublon bound state applies to the anomalous Zeeman shift of the antispinon holon bound states as well.

%The Zeeman field effect on the energy saving in the binding is further verified by the binding susceptibility calculation.

Below we discuss more quantitatively how this anomalous Zeeman shift comes about. Let us take the electronic bound state below the UHB as an example and the case of hole bound state above the LHB can be found in the Supplemental Materials~\cite{Supp}. We define the susceptibility to bind a spinon and a doublon, which is the real part of the bare electronic Green's function obtained by convolving the spinon and doublon Green's function~\cite{Supp}
\begin{align}
\chi_{d,\sigma}\left(\bm{k},\omega\right)=&-\frac{1}{N}\sum_{\bm{q}} \textrm{Re} G^{\textrm{R}}_{d,\sigma}\left(\omega,\bm{k}-\bm{q},\bm{q}\right)
\end{align}
with $\bm{k}_{\textrm{F}}$ being the spinon Fermi wave vector. {Here the form of the bare Green's function $G^{\textrm{R}}_{d,\sigma}\left(\omega,\bm{k}-\bm{q},\bm{q}\right)$, which involves no gauge binding, is given in Eq. S9 in the Supplemental Materials~\cite{Supp}.} In Fig. \ref{figure3} (b), $\chi_{d,\uparrow}\left(\bm{k}_{\textrm{F}},\omega\right)$ is plotted given different Zeeman fields. We see that the susceptibility $\chi_{d,\uparrow}\left(\bm{k}_{\textrm{F}},\omega\right)$ is larger for the spinons of parallel spin compared with those of anti-parallel spin. When the gauge binding is sufficiently large, the bound state is solved from the binding equation $U_{\textrm{b}}\chi_{d,\sigma}\left(\bm{d},\omega\right)=1$~\cite{Supp}. This is solved graphically by finding the intercept of $\chi_{d,\uparrow}\left(\bm{k}_{\textrm{F}},\omega\right)$ plotted in Fig. \ref{figure3} (b) with $1/U_{\textrm{b}}$. For sufficiently large $U_{\textrm{b}}$, the intercept at $\omega<\Delta$ gives the bound state energy. It is clear  that the intercept gets a smaller value for the spin parallel to the magnetic field.

%spinexhibit the anomalous Zeeman shift: the state of anti-parallel spin lowers the energy while that of parallel spin has the energy lifted. In the strong gauge binding regime, the Zeeman field effect on the energy saving dominates over the spinon Zeeman energy $\mu_{\textrm{b}}B$ and therefore brings about the anomalous Zeeman shift of the bound states.

\section{Conclusions and discussions}
In the above sections, the QSL electronic DOS spectra in a Zeeman field is found to exhibit zero shift for the threshold energy, naive shift for the quasi-bound spinon chargon pairs, and anomalous shift for the in-gap bound states. In experiments which inject unpolarized electrons, the Zeeman splitting of the band edge resonance or the bound state will be smeared out by the thermal fluctuations. Therefore to see this effect it will be best to inject electron with one kind of spin via spin polarized STM~\cite{Bode,Balatsky}. In order to suppress the thermal smearing, it is better to carry out the measurement in the low temperature $k_{\textrm{b}}T\ll\mu_{\textrm{b}}B$, where $k_{\textrm{b}}$ is the Boltzman constant. In practice, one would focus on measuring the threshold, Hubbard band widths and the band edge resonance peaks. Observation of the zero Zeeman shift of the threshold, Zeeman field induced change of the Hubbard band widths, and naive Zeeman shift of the band edge resonance peaks if any are evidence supporting the existence of the SFS. If the electronic DOS spectra as a whole exhibits the anomalous Zeeman shift, then our study indicates that the spectra is composed of the bounded spinon chargon pairs. However, it is difficult to realize this limit, because in this case the bound state will exhaust a large fraction of the spectral weight, as seen in Fig. \ref{figure2} (c). {The local electronic DOS spectrum obtained by the STM measurements~\cite{Yayu, Yichen2, Vano, WeiRuan, Butler1, Butler2} so far shows either no resonance or resonaces with relatively small spectral weight, therefore favoring the zero gauge binding case in Fig. \ref{figure1} (c) or the weak gauge binding induced band edge resonance scenario in Fig. \ref{figure2} (b).} It is also important to note that the zero Zeeman shift of the threshold energy serves to distinguish between the QSL with SFS and a gapped QSL which is expected to show a conventional Zeeman shift.

Usually, the effect of magnetic field on materials is two fold: one is the Zeeman effect and the other is the orbital effect. {The orbital effect of magnetic field on the QSL with SFS is detailed studied in our companion paper~\cite{Wenyu3}. The effect of orbital magnetic field is to Landau quantize the QSL with SFS, and the Landu quantization feature of the QSL is characterized by the Hubbard band edge steps and resonance peaks in the electronci DOS~\cite{Wenyu3}.} Fortunatley, owing to the recent emerging two-dimensional Mott insulator 1T-TaSe$_2$~\cite{Yichen, Yichen2, WeiRuan}, 1T-NbSe$_2$~\cite{Takahashi, Mengke, Yeliang, YingShuang} and the surface of the layered 1T-TaS$_2$~\cite{Butler1, Butler2, Shichao}, one can apply an in-plane magnetic field as shown in Fig. \ref{figure3} (a) to get rid of the orbital magnetic field effect~\cite{Wenyu3}. Therefore the two-dimensional QSL candidate materials with an in-plane magnetic field are ideal platforms to study the QSL electronic Zeeman response. Recently, a clear UHB threshold along with a UHB edge resonance peak has been observed in the STM spectra on the surface of the layered 1T-TaS$_2$~\cite{Butler2}. From our study, a further spin polarized STM measurement on the evolution of the DOS spectra with an in-plane magnetic field would be  helpful to identify the ground state on the surface of 1T-TaS$_2$. In many cases, strong in-plane magnetic field is not available for STM experiments. In that case it will be instructive to compare spin injection from STM tip with opposite spin polarization to isolate the effect of the Zeeman field from the orbital effect~\cite{Jinfeng}.

\section*{ACKNOWLEDGEMENT}
W.-Y. He acknowledges the start-up grant of ShanghaiTech University. P. A. Lee acknowledges support by DOE office of Basic Sciences grant number DE-FG02-03ER46076.

\onecolumngrid
\clearpage
\begin{center}

{\bf Supplemental Material for ``Electronic Density of States of a $U\left(1\right)$ Quantum Spin Liquid with Spinon Fermi Surface. II. Zeeman Magnetic Field Effects''}

\end{center}

\maketitle
\setcounter{equation}{0}
\setcounter{figure}{0}
\setcounter{table}{0}
\setcounter{page}{1}
\makeatletter

\renewcommand{\theequation}{S\arabic{equation}}
\renewcommand{\thefigure}{S\arabic{figure}}
\renewcommand{\thetable}{S\arabic{table}}
\renewcommand{\bibnumfmt}[1]{[S#1]}
\renewcommand{\citenumfont}[1]{S#1}

\section{Electronic Green's function in the Quantum Spin Liquid}
Following the slave rotor mean field formalism~\cite{Florens_S, Patrick}, one can derive the action for the $U\left(1\right)$ quantum spin liquid with spinon Fermi sea~\cite{WenyuKondo}
\begin{align}\nonumber\label{Action_0}
S=&-\sum_{\sigma,\bm{k},\omega_n}\left(i\omega_n-\xi_{\sigma,\bm{k}}\right)f^\dagger_{\sigma,\bm{k},n}f_{\sigma,\bm{k},n}-\sum_{\bm{k}}\left(-i\nu_n-\epsilon_{\bm{k}}\right)a_{-\bm{k},n}a^\dagger_{-\bm{k},n}-\sum_{\bm{k}}\left(i\nu_n-\epsilon_{\bm{k}}\right)b^\dagger_{\bm{k},n}b_{\bm{k},n}\\
&+\frac{U_{\textrm{b}}}{N}\sum_{\sigma,\bm{k},\bm{q},\bm{q}'}f^\dagger_{\sigma,\bm{k}-\bm{q}}f_{\sigma,\bm{k}-\bm{q}'}\left(a_{-\bm{q}}a^\dagger_{-\bm{q}'}-b^\dagger_{\bm{q}}b_{\bm{q}'}\right),
\end{align}
where $a^{\left(\dagger\right)}_{-\bm{k}}$, $b^{\left(\dagger\right)}_{\bm{k}}$ and $f_{\sigma,\bm{k}}^{\left(\dagger\right)}$ are the annihilation (creation) operators for a holon, doublon and spinon respectively. Here $U_{\textrm{b}}$ denotes the gauge binding arising from the longitudinal gauge field fluctuations, $\xi_{\sigma,\bm{k}}$ is the spinon band dispersion, $\epsilon_{\bm{k}}$ is the band dispersion of the holon and doublon. The relativistic slave rotor has been approximated by the relativistic holon and doublon, which captures the low energy physics near the Hubbard band edges.

In the $U\left(1\right)$ quantum spin liquid with spinon Fermi sea, creationg an electron is to create a spinon and simultaneously create a doublon or annihilate a holon. The creation operator for an electronic state is $c^\dagger_{\bm{k},\bm{k}',\sigma}=f^\dagger_{\sigma,\bm{k}}\left(a_{-\bm{k}'}+b^\dagger_{\bm{k}'}\right)$. In the absence of gauge binding, the bare electronic Matsubara Green's function at $U_{\textrm{b}}=0$ takes the form
\begin{align}\nonumber\label{Green_electronic}
G_\sigma\left(i\omega_n,\bm{k},\bm{k}'\right)=&-\int_0^\beta\left\langle c_{\bm{k},\bm{k}',\sigma}\left(\tau\right)c^\dagger_{\bm{k},\bm{k}',\sigma}\left(0\right) \right\rangle_0e^{i\omega_n\tau} d\tau\\\nonumber
=&-\int_0^\beta\left\langle f_{\sigma,\bm{k}}\left(\tau\right)f^\dagger_{\sigma,\bm{k}}\left(0\right) \right\rangle_0\left[\left\langle a^\dagger_{-\bm{k}'}\left(\tau\right)a_{-\bm{k}'}\left(0\right) \right\rangle_0+\left\langle b_{\bm{k}'}\left(\tau\right)b^\dagger_{\bm{k}'}\left(0\right) \right\rangle_0\right]e^{i\omega_n\tau}d\tau\\
=&-\frac{1}{\beta}\sum_{\nu_n}G_{f,\sigma}\left(i\omega_n-i\nu_n,\bm{k}\right)\left[G_a\left(-i\nu_n,-\bm{k}'\right)+G_b\left(i\nu_n,\bm{k}'\right)\right],
\end{align}
where $G_{f,\sigma}\left(i\omega_n,\bm{k}\right)$ is the spinon Matsubara Green's function
\begin{align}
G_{f,\sigma}\left(i\omega_n,\bm{k}\right)=-\int_0^\beta\left\langle f_{\sigma,\bm{k}}\left(\tau\right)f^\dagger_{\sigma,\bm{k}}\left(0\right) \right\rangle_0e^{i\omega_n\tau}d\tau=\frac{1}{i\omega_n-\xi_{\bm{k}}},
\end{align}
$G_a\left(-i\nu_n,-\bm{k}\right)$ is the holon Matsubara Green's function
\begin{align}
G_a\left(-i\nu_n,-\bm{k}\right)=&-\int_0^\beta\left\langle a^\dagger_{-\bm{k}}\left(\tau\right)a_{-\bm{k}}\left(0\right) \right\rangle_0e^{-i\nu_b\tau}d\tau=\frac{1}{-i\nu_n-\epsilon_{\bm{k}}},
\end{align}
and $G_b\left(i\nu_n,\bm{k}\right)$ is the doublon Matsubara Green's function
\begin{align}
G_b\left(i\nu_n,\bm{k}\right)=&-\int_0^\beta\left\langle b_{\bm{k}}\left(0\right)b^\dagger_{\bm{k}}\left(0\right) \right\rangle_0e^{i\nu_n\tau}d\tau=\frac{1}{i\nu_n-\epsilon_{\bm{k}}}.
\end{align}
Here $\left\langle \dots \right\rangle_0$ denotes the thermal average calculated from $S$ in Eq. \ref{Action_0} at zero gauge binding. After performing the Matsubara frequency summation, one can obtain the form of the electronic Matasubara Green's function in Eq. \ref{Green_electronic}
\begin{align}
G_\sigma\left(i\omega_n,\bm{k},\bm{k}'\right)=&G_{h,\sigma}\left(i\omega_n,\bm{k},\bm{k}'\right)+G_{d,\sigma}\left(i\omega_n,\bm{k},\bm{k}'\right)
\end{align}
with
\begin{align}
G_{h,\sigma}\left(i\omega_n,\bm{k},\bm{k}'\right)=\frac{n_{\textrm{F}}\left(\xi_{\sigma,\bm{k}}\right)+n_{\textrm{B}}\left(\epsilon_{\bm{k}'}\right)}{i\omega_n-\xi_{\sigma,\bm{k}}+\epsilon_{\bm{k}'}},\quad G_{d,\sigma}\left(i\omega_n,\bm{k},\bm{k}'\right)=\frac{n_{\textrm{F}}\left(-\xi_{\sigma,\bm{k}}\right)+n_{\textrm{B}}\left(\epsilon_{\bm{k}'}\right)}{i\omega_n-\xi_{\sigma,\bm{k}}-\epsilon_{\bm{k}'}}.
\end{align}
The corresponding retarded Green's function can then be obtained through the analytic continuation $i\omega_n\rightarrow\omega+i0^+$
\begin{align}
G_{\sigma}^{\textrm{R}}\left(\omega,\bm{k},\bm{k}'\right)=&G^{\textrm{R}}_{h,\sigma}\left(\omega,\bm{k},\bm{k}'\right)+G^{\textrm{R}}_{d,\sigma}\left(\omega,\bm{k},\bm{k}'\right)
\end{align}
with
\begin{align}
G^{\textrm{R}}_{h,\sigma}\left(\omega,\bm{k},\bm{k}'\right)=\frac{n_{\textrm{F}}\left(\xi_{\sigma,\bm{k}}\right)+n_{\textrm{B}}\left(\epsilon_{\bm{k}'}\right)}{\omega+i0^+-\xi_{\sigma,\bm{k}}+\epsilon_{\bm{k}'}},\quad G^{\textrm{R}}_{d,\sigma}\left(\omega,\bm{k},\bm{k}'\right)=\frac{n_{\textrm{F}}\left(-\xi_{\sigma,\bm{k}}\right)+n_{\textrm{B}}\left(\epsilon_{\bm{k}'}\right)}{\omega+i0^+-\xi_{\sigma,\bm{k}}-\epsilon_{\bm{k}'}}.
\end{align}
Here the first term and second term in $G_{\sigma}^{\textrm{R}}\left(\omega,\bm{k},\bm{k}'\right)$ describe the states in the lower Hubbard band and upper Hubbard band respectively.

\section{Electronic	Density of States in the Quantum Spin Liquid}
The electronic density of states at the site $\bm{r}_i$ takes the form
\begin{align}
\tilde{\rho}_\sigma\left(\omega\right)=&-\frac{1}{\pi}\textrm{Im}\tilde{G}^{\textrm{R}}_\sigma\left(\bm{r}_i,\bm{r}_i,\omega\right),
\end{align}
where $\tilde{G}^{\textrm{R}}_\sigma\left(\bm{r}_i,\bm{r}_i,\omega\right)$ is the electronic retarded Green's function in real space. The corresponding electronic Matsubara Green's function in real space takes the form
\begin{align}
\tilde{G}_\sigma\left(\bm{r}_i,\bm{r}_j,i\omega_n\right)=&-\int_0^\beta\left[\left\langle f_{i,\sigma}\left(\tau\right)f^\dagger_{j,\sigma}\left(0\right)a^\dagger_i\left(\tau\right)a_j\left(0\right) \right\rangle+\left\langle f_{i,\sigma}\left(\tau\right)f^\dagger_{j,\sigma}\left(0\right)b_i\left(\tau\right)b_j^\dagger\left(0\right) \right\rangle\right]e^{i\omega_n\tau}d\tau
\end{align}
where $\left\langle \dots \right\rangle$ denotes the thermal average calculated from $S$ in Eq. \ref{Action_0} that involves the gauge binding $U_{\textrm{b}}$. In the system with translational symmetry, we can perform the Fourier transformation and then get
\begin{align}\nonumber
\tilde{G}_\sigma\left(i\omega_n,\bm{k}\right)=&\sum_{\delta\bm{r}}\tilde{G}_\sigma\left(\bm{r}_i,\bm{r}_j,i\omega_n\right)e^{-i\bm{k}\cdot\left(\bm{r}_i-\bm{r}_j\right)}\\\nonumber
=&-\sum_{\delta\bm{r}}\int_0^\beta\left\langle f_{i,\sigma}\left(\tau\right)f^\dagger_{j,\sigma}\left(0\right)a^\dagger_i\left(\tau\right)a_j\left(0\right) \right\rangle e^{i\omega_n\tau-i\bm{k}\cdot\left(\bm{r}_i-\bm{r}_j\right)}d\tau\\
&-\sum_{\delta\bm{r}}\int_0^\beta\left\langle f_{i,\sigma}\left(\tau\right)f^\dagger_{j,\sigma}\left(0\right)b_i\left(\tau\right)b_j^\dagger\left(0\right) \right\rangle e^{i\omega_n\tau-i\bm{k}\cdot\left(\bm{r}_i-\bm{r}_j\right)}d\tau
\end{align}
The first term can be further simplified as
\begin{align}\nonumber\label{Gh01}
-\sum_{\delta\bm{r}}\int_0^\beta\left\langle f_{i,\sigma}\left(\tau\right)f^\dagger_{j,\sigma}\left(0\right)a^\dagger_i\left(\tau\right)a_j\left(0\right) \right\rangle e^{i\omega_n\tau-i\bm{k}\cdot\left(\bm{r}_i-\bm{r}_j\right)}d\tau=&-\frac{1}{N}\sum_{\bm{q}}\int_0^\beta\left\langle f_{\sigma,\bm{k}-\bm{q}}\left(\tau\right)f^\dagger_{\sigma,\bm{k}-\bm{q}}\left(0\right)a^\dagger_{-\bm{q}}\left(\tau\right)a_{-\bm{q}}\left(0\right)\right\rangle e^{i\omega_n\tau}d\tau\\
=&\frac{\frac{1}{N}\sum_{\bm{q}}G_{h,\sigma}\left(i\omega_n,\bm{k}-\bm{q},\bm{q}\right)}{1-\frac{U_{\textrm{b}}}{N}\sum_{\bm{q}}G_{h,\sigma}\left(i\omega_n,\bm{k}-\bm{q},\bm{q}\right)},
\end{align}
and similarly the second term is also simplified as
\begin{align}\nonumber\label{Gd01}
-\sum_{\delta\bm{r}}\int_0^\beta\left\langle f_{i,\sigma}\left(\tau\right)f^\dagger_{j,\sigma}\left(0\right)b_i\left(\tau\right)b_j^\dagger\left(0\right) \right\rangle e^{i\omega_n\tau-i\bm{k}\cdot\left(\bm{r}_i-\bm{r}_j\right)}d\tau=&-\frac{1}{N}\sum_{\bm{q}}\int_0^\beta\left\langle f_{\sigma,\bm{k}-\bm{q}}\left(\tau\right)f^\dagger_{\sigma,\bm{k}-\bm{q}}\left(0\right)b_{\bm{q}}\left(\tau\right)b^\dagger_{\bm{q}}\left(0\right)\right\rangle e^{i\omega_n\tau}d\tau\\
=&\frac{\frac{1}{N}\sum_{\bm{q}}G_{d,\sigma}\left(i\omega_n,\bm{k}-\bm{q},\bm{q}\right)}{1+\frac{U_{\textrm{b}}}{N}\sum_{\bm{q}}G_{d,\sigma}\left(i\omega_n,\bm{k}-\bm{q},\bm{q}\right)}.
\end{align}
Here we have used the random approximation in the second step in Eq. \ref{Gh01} and \ref{Gd01}. By applying the analytic continuation $i\omega_n\rightarrow\omega+i0^+$, we eventually get the electronic density of states $\tilde{\rho}_\sigma\left(\omega\right)$ to be
\begin{align}\nonumber\label{electronic_DOS}
\tilde{\rho}_\sigma\left(\omega\right)=&-\frac{1}{\Omega\pi}\sum_{\bm{k}}\textrm{Im}\tilde{G}_\sigma^{\textrm{R}}\left(\omega,\bm{k}\right)\\
=&-\frac{1}{\Omega\pi}\sum_{\bm{k}}\textrm{Im}\frac{\frac{1}{N}\sum_{\bm{q}}G^{\textrm{R}}_{h,\sigma}\left(\omega,\bm{k}-\bm{q},\bm{k}\right)}{1-\frac{U_{\textrm{b}}}{N}\sum_{\bm{q}}G^{\textrm{R}}_{h,\sigma}\left(\omega,\bm{k}-\bm{q},\bm{k}\right)}-\frac{1}{\Omega\pi}\sum_{\bm{k}}\textrm{Im}\frac{\frac{1}{N}\sum_{\bm{q}}G^{\textrm{R}}_{d,\sigma}\left(\omega,\bm{k}-\bm{q},\bm{q}\right)}{1+\frac{U_{\textrm{b}}}{N}\sum_{\bm{q}}G^{\textrm{R}}_{d,\sigma}\left(\omega,\bm{k}-\bm{q},\bm{q}\right)}.
\end{align}

\section{The Binding Susceptibilities and Binding Equations}
In the presence of gauge binding, one can see from Eq. \ref{electronic_DOS} that the gauge binding $U_{\textrm{b}}$ changes the electronic density of states drastically when $1-\frac{U_{\textrm{b}}}{N}\sum_{\bm{q}}G^{\textrm{R}}_{h,\sigma}\left(\omega,\bm{k}-\bm{q},\bm{q}\right)\rightarrow0$ and $1+\frac{U_{\textrm{b}}}{N}\sum_{\bm{q}}G^{\textrm{R}}_{d,\sigma}\left(\omega,\bm{k}-\bm{q},\bm{q}\right)\rightarrow0$. In the regime of weak gauge binding, the first term in Eq. \ref{electronic_DOS} gives rise to a resonance peak at the LHB top while the second term in Eq. \ref{electronic_DOS} brings about a resonance peak at the UHB bottom. When the gauge binding is sufficiently large, the band edge resonane peaks evolve inside the Mott gap and develop into in-gap bound states. The critical conditions $1-\frac{U_{\textrm{b}}}{N}\sum_{\bm{q}}G^{\textrm{R}}_{h,\sigma}\left(\omega,\bm{k}-\bm{q},\bm{q}\right)=0$ and $1+\frac{U_{\textrm{b}}}{N}\sum_{\bm{q}}G^{\textrm{R}}_{d,\sigma}\left(\omega,\bm{k}-\bm{q},\bm{q}\right)=0$ give the binding equations that determine the bound state band dispersions:
\begin{align}\label{binding_Eq1}
U_{\textrm{b}}\chi_{h,\sigma}\left(\bm{k},\omega\right)=&1,\\\label{binding_Eq2}
U_{\textrm{b}}\chi_{d,\sigma}\left(\bm{k},\omega\right)=&1.
\end{align}
Here Eq. \ref{binding_Eq1} gives the band dispersion for the hole bound state above the LHB while Eq. \ref{binding_Eq2} gives the band dispersion for the electronic bound state below the UHB. In Eq. \ref{binding_Eq1} and Eq. \ref{binding_Eq2}, $\chi_{h,\sigma}\left(\bm{k},\omega\right)$ and $\chi_{d,\sigma}\left(\bm{k},\omega\right)$ are the susceptibilities to form a hole bound state and an electronic bound state respectively. The two susceptibilities take the form
\begin{align}
\chi_{h,\sigma}\left(\bm{k},\omega\right)=&\frac{1}{N}\sum_{\bm{q}}\textrm{Re}G^{\textrm{R}}_{h,\sigma}\left(\omega,\bm{k}-\bm{q},\bm{q}\right),\quad\textrm{and}\quad\chi_{d,\sigma}\left(\bm{k},\omega\right)=-\frac{1}{N}\sum_{\bm{q}}\textrm{Re}G^{\textrm{R}}_{d,\sigma}\left(\omega,\bm{k}-\bm{q},\bm{q}\right)
\end{align}
respectively.

\end{document}